\nonstopmode
\documentclass{article}

\usepackage{fixltx2e}
\usepackage{amssymb}

\usepackage{amsmath}

\usepackage{graphicx}

\usepackage{tikz}

\usepackage{subcaption}

\usepackage{geometry}
\usepackage{hyperref}

\date{2010 Dec 4 (revised 2014 Dec 28)}

\author{R.~P.~Munafo \footnote{\small 10 Linwood St. \#304, Malden MA 02148, United States; mrob@mrob.com; http://mrob.com/sci}}

\title{Stable localized moving patterns in the 2-D Gray-Scott model}

\begin{document}

\maketitle

\hrule

\def\dlu/{\textsf{lu}}
\def\dtu/{\textsf{tu}}
\def\typeB/{\textbf{\textsf{B}}}
\def\typeL/{\textbf{\textsf{L}}}
\def\typeR/{\textbf{\textsf{R}}}

\setlength{\unitlength}{1.0cm}

\begin{abstract}

I show stable, localized, single and multi-spot patterns of three classes --
stationary, moving, and rotating -- that exist within a limited range
of parameter values in the two-dimensional Gray-Scott reaction-diffusion
model with $\sigma=D_u/D_v=2$. These patterns exist in domains of any
size, and appear to derive their stability from a constructive
reinforcement effect of the standing waves that surround any feature.
There are several common elements --
including a spot that behaves as a quasiparticle, a U-shaped
stripe, and a ring or annulus, or a portion thereof --
which combine to form a great variety of stable structures.
These patterns interact with each other in
a variety of ways. There are similarities to
other reaction-diffusion systems and to physical experiments; I
offer several suggestions for further research.



\end{abstract}

\large

\section{Introduction}
\label{intro}

The Gray-Scott model \cite{gray1983} is a widely-studied model of a
pair of reactions involving cubic autocatalysis. It has been applied
in reaction-diffusion models in one \cite{mazin1996, doelman1997,
muratov2000, nishiura2000, kolokolnikov2004},
two \cite{pearson1993, muratov2000, wei2003, mcgough2004, munteanu2007}
and three \cite{leppanen2004, lidbeck2007} dimensions. Widely-known results
include the existence of stable single spots, self-replication of
spots, spontaneous formation of stripes and hexagonal arrays of spots,
and Turing patterns \cite{mazin1996}. Findings that show a lack of
moving stable patterns, e.g. \cite{doelman1997} apply only in 1-D or a
limited region of the 2-D system parameter space (\cite{muratov2000}
p. 81 and \cite{wei2003} p. 3). \cite{wei2003} shows existence of
stable multi-spot patterns within a finite domain.

The model equations are:
\begin{equation} \label{uv-pde}
  \begin{aligned}
    \frac{\partial u}{\partial t} &= D_u\nabla^2u - uv^2 + F(1 - u) , \\
    \frac{\partial v}{\partial t} &= D_v\nabla^2v + uv^2 - (F + k)v .
  \end{aligned}
\end{equation}

This paper concerns the two-dimensional case with periodic boundary
conditions. I use similar terminology and symbols to those in
\cite{pearson1993}: $u$ and $v$ are the concentrations of two
reactants $U$ and $V$, normalized to dimensionless units. Parameters
$F$ and $k$ represent the feed rate and removal rate of the reactants
in the original homogeneous continuously-stirred tank reactor model;
in diffusion systems they are typically the rates at which $U$ and $V$
permeate through a membrane separating a homogeneous supply from the
gel in which the patterned reactions occur \cite{lee1993}. $D_u$ and
$D_v$ are the diffusion rates of the two reactants through the gel or
other medium; considered together as a ratio $\sigma = D_u/D_v$, they
constitute a third parameter that determines certain characteristics
of the parameter space \cite{mazin1996}. In this paper, $D_u = 2
\times 10^{-5}$ and $D_v = 10^{-5}$. Since I am measuring the size,
velocity, and other related statistics of certain features in observed
patterns, there is a need for time and length units. For this purpose
I use the dimensionless units of length and time implicit in the
equations, and refer to them as \dlu/ (``length unit'') and \dtu/
(``time unit'') respectively.

\section{Methods}

I used numerical simulation similar to that of Pearson
\cite{pearson1993} (discrete Euler forward integration, also called
forward-time centered-space \cite{press1992}). All figures shown here
were produced with a $256 \times 256$ grid, with $\Delta x =
1/143~\dlu/$ representing a system size of $1.79~\dlu/
\times~1.79~\dlu/$, periodic boundary conditions, and $\Delta t =
1/2~\dtu/$.

Results were verified using higher grid resolution, smaller time
steps, and 2\textsuperscript{nd} and 4\textsuperscript{th} order Runge-Kutta integration. Because
many of these patterns exhibit novel properties (such as stability
combined with motion) it was necessary to perform a thorough series of
measurements to isolate the legitimate ``real'' phenomena from any
effects that arise from simulation error.

In these tests, the grid resolution $\Delta x$, time step $\Delta t$,
and stability index $\Delta t/(\Delta x)^2$ (the variable component of
the Courant condition, see \cite{press1992, mazin1996}) were each
decreased in steps of $\sqrt{2}$, $2\sqrt{2}$ and $\sqrt{2}$
respectively. The finest resolution used was $\Delta x = 1/572~\dlu/,
\Delta t = 1/128~\dtu/$ with double-precision arithmetic.

\vspace{\baselineskip}
\hrule

\small

\begin{table}[h]
 \begin{center}
  \begin{tabular}{c c c c c c}
    &  &  & \multicolumn{2}{c}{measured velocities}
                                       & range of stable \textsf{U} pattern \\
 $\Delta x$ & $\Delta t$ & $(\Delta x)^2/\Delta t$  & \textsf{U} & 3-dot
                                            & (to within $\pm 10^{-7}$) \\
 \hline
 $1/143$         & $1/2$
             & $9.78 \times 10^{-5}$ & $1~\dlu//62277~\dtu/$
                                         & $1~\dlu//9.1211 \times 10^6~\dtu/$
                                         & $0.0608833 \le k \le 0.0609829$ \\
 $1/143\sqrt{2}$ & $1/4\sqrt{2}$
             & $1.38 \times 10^{-4}$ & $1~\dlu//62204~\dtu/$
                                         & $1~\dlu//9.0037 \times 10^6~\dtu/$
                                         & $0.0608796 \le k \le 0.0609831$ \\
 $1/286$         & $1/16$
             & $1.96 \times 10^{-4}$ & $1~\dlu//62300~\dtu/$
                                         & $1~\dlu//8.8740 \times 10^6~\dtu/$
                                         & $0.0608778 \le k \le 0.0609832$ \\
 $1/286\sqrt{2}$ & $1/32\sqrt{2}$
             & $2.77 \times 10^{-4}$ & $1~\dlu//62161~\dtu/$
                                                     & $d.n.t.$ & $d.n.t.$ \\
 $1/572$         & $1/128$
             & $3.91 \times 10^{-4}$ & $1~\dlu//62012~\dtu/$
                                                     & $d.n.t.$ & $d.n.t.$ \\
  \end{tabular}
  \hspace*{2cm}\caption{Measurements of the velocities of the two moving patterns
  in figure~\ref{fig:pi-valu}, performed with progressively increasing
  grid and time resolutions. All tests were performed in IEEE double
  precision; $F = 0.06$, $k = 0.0609$, and gridsize $0.9~\dlu/ \times
  0.9~\dlu/$ or larger. \endgraf\setlength{\parindent}{1em}
  Procedure for velocities: 1. Wait for moving
  pattern to assume its steady-state appearance. 2. Locate a grid
  point on the centerline of the moving feature and on its leading
  edge, where $\partial u/\partial t$ is high and $u$ is close to
  0.42; note its location and the elapsed system time (in \dlu/ and
  \dtu/ respectively). 3. Wait for the pattern to travel at least $200
  \Delta x$. 4. Measure again, compute $v = d/t$. Measurement error is
  1 part in 200. \endgraf
  Procedure for last column: 1. $F = 0.06$; $k$ set to trial value;
  grid contains a single \textsf{U} pattern. 2. At periodic intervals,
  compute sum of all $u$ values in the grid. 3. Examine the $u$ sums to
  identify its asymptotic behavior: a stable moving \textsf{U} approaches a
  finite asymptote and continues moving; unstable patterns stop moving
  or grow without bound. 4. Repeat for a new $k$ value; continue until
  the range of valid $k$ values is known to the desired precision. \endgraf
  $d.n.t.$ = Did not test.
  \label{tab:accuracy}
 }\hspace*{2cm}
 \end{center}
\end{table}

\begin{table}[h]
 \begin{center}
  \begin{tabular}{c c c c c c}
    &  &  & \multicolumn{2}{c}{measured velocities}
                                       & range of stable \textsf{U} pattern \\
 $\Delta x$ & $\Delta t$ & $(\Delta x)^2/\Delta t$  & \textsf{U} & 3-dot
                                            & (to within $\pm 10^{-7}$) \\
 \hline
 $1/143$         & $1/2$
             & $9.78 \times 10^{-5}$ & $1~\dlu//62420~\dtu/$
                                         & $1~\dlu//9.1911 \times 10^6~\dtu/$
                                         & $0.0608833 \le k \le 0.0609829$ \\
 $1/143\sqrt{2}$ & $1/4\sqrt{2}$
             & $1.38 \times 10^{-4}$ & $1~\dlu//62477~\dtu/$
                                         & $1~\dlu//9.0987 \times 10^6~\dtu/$
                                         & $0.0608796 \le k \le 0.0609831$ \\
 $1/286$         & $1/16$
             & $1.96 \times 10^{-4}$ & $1~\dlu//62239~\dtu/$
                                         & $1~\dlu//1.0722 \times 10^7~\dtu/$
                                         & $0.0608777 \le k \le 0.0609831$ \\
 $1/286\sqrt{2}$ & $1/32\sqrt{2}$
             & $2.77 \times 10^{-4}$ & $1~\dlu//62744~\dtu/$
                                                     & $d.n.t.$ & $d.n.t.$ \\
 $1/572$         & $1/128$
             & $3.91 \times 10^{-4}$ & $1~\dlu//62571~\dtu/$
                                                     & $d.n.t.$ & $d.n.t.$ \\
  \end{tabular}
  \caption{The same tests as in Table 1, performed in IEEE single precision floating-point.
  \label{tab:single}
  }
 \end{center}
\end{table}

\vspace{\baselineskip}
\large

These tests showed no qualitative difference in the results, but did
enable more accurate determination of certain measurements. For
example, the ``measured velocities'' in Tables 1 and 2 have a
measurement error of 1 part in 200 due to the procedure used, and the
increases in $\Delta x$ and $\Delta t$ resolution had no significant
effect on the measured velocities: but the double-precision tests show
a somewhat greater consistency (i.e. lower variance) in the same
measurement across different grid resolutions.

All phenomena reported in the remainder of this paper appear at the
grid resolutions of \cite{pearson1993}, $\Delta x = 1/102~\dlu/, \Delta
t = 1~\dtu/$, with single-precision arithmetic. A few grid effects are
apparent at this lower resolution, for example the two-spot pattern in
figure~\ref{fig:pi-valu} reorients to a diagonal alignment. Such
effects diminish sharply at higher grid resolutions.

\begin{figure}
  \centering
  \begin{subfigure}[b]{0.2\textwidth}
    \includegraphics[width=\textwidth]{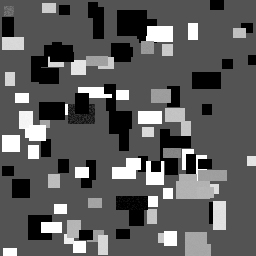}
    \caption{start}
    \label{fig:uskate-seq-t00000.jpg}
  \end{subfigure}
  \hspace{2mm}
  \begin{subfigure}[b]{0.2\textwidth}
    \includegraphics[width=\textwidth]{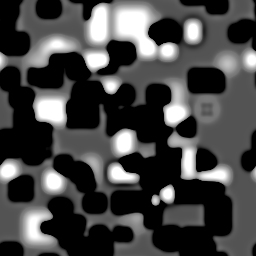}
    \caption{$t = 32~\dtu/$}
    \label{fig:uskate-seq-t00032.jpg}
  \end{subfigure}
  \hspace{2mm}
  \begin{subfigure}[b]{0.2\textwidth}
    \includegraphics[width=\textwidth]{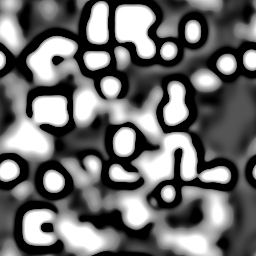}
    \caption{$t = 120~\dtu/$}
    \label{fig:uskate-seq-t00120.jpg}
  \end{subfigure}
  \hspace{2mm}
  \begin{subfigure}[b]{0.2\textwidth}
    \includegraphics[width=\textwidth]{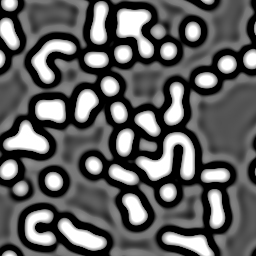}
    \caption{$t = 480~\dtu/$}
    \label{fig:uskate-seq-t00480.jpg}
  \end{subfigure}
  \caption{
    Starting pattern used to generate
    figure~\ref{fig:wild-pi}. In the first
    image, each rectangle has a randomly chosen $u$ and $v$ between
    0 and 1, and the background is $(u, v) \approx (0.4201, 0.2878)$.
    Coloring is the same as in figure~\ref{fig:types-nx}. The
    exaggerated contrast makes many $u$ values completely black or white.
    \label{fig:uskate-generation}
  }
\end{figure}

The initial state was created in several ways. Most results were
produced by starting with a background level of $u$ and $v$ set to a
homogeneous state computed from $F$ and $k$. There is always a stable
state $(u_{h1}=1, v_{h1}=0)$, sometimes called the ``red state''
\cite{pearson1993, lee1994, mazin1996}. For $(F, k)$ sufficiently
small there are two other homogeneous states, one of which can be
stable (called the ``blue state'', due to the color of
the pH indicator bromothymol blue the laboratory experiments of \cite{lee1993}).
Transforming the variables and units from Muratov and
Osipov\cite{muratov2000}, the third homogeneous state $(u_{h3},
v_{h3})$ exists when $k < (\sqrt{F}-2F)/2$, with $u$ and $v$ given by
\begin{align} \label{uv-h3}
  u_{h3} &= \frac{A - \sqrt{A^2-4}}{2A} , &
  v_{h3} &= \frac{\sqrt{F}(A + \sqrt{A^2 - 4})}{2}
\end{align}
where $A$ is $\sqrt{F}/(F+k)$. Starting with a grid filled with these
values of $u$ and $v$ or with $(u=0, v=1)$, a number of rectangles
were added whose width, height, locations, and number were determined
randomly, and then filled with random levels of $u$ and $v$. For most
tests, these rectangles ranged from $5 \times 10^{-3}~\dlu/^2$ to $4
\times 10^{-2}~\dlu/^2$ in area, and the frequency of occurrence of any
given size rectangle was inversely proportional to its area. Density
of these rectangles ranged anywhere from 1 to 40 per $\dlu/^2$. A
typical example is shown in figure~\ref{fig:uskate-generation}.

A repeatable pseudo-random algorithm was used to generate all starting
patterns, and the initial seed values saved so the same simulation
sequence could be reproduced at will \cite{knuth1997}. The simulation
was run for anywhere from $10^5$ to $5 \times 10^8$ \dtu/, as needed
for the phenomena under test. A complete survey of the parameter space
in figure~\ref{fig:param-space} was performed at intervals of 0.02 in
$k$ and 0.04 in $F$; many starting patterns were tried at each
parameter value combination. The interested reader can find a gallery
of images online at \cite{munafo2009gallery}.

Although all of the patterns shown here arose naturally from such
initial random states, a number of techniques were used to make
exploration and discovery more practical: Selectively removing
unwanted patterns by setting portions of the grid to the homogeneous
state; combining parts of patterns to create others; changing
parameter values; moving small parts of the grid to test pattern
integrity; then continuing the simulation after making any of these
changes.

Because of their stability, each type $\pi$ pattern naturally evolves
from any starting pattern that superficially resembles it, provided
the dimensions and levels of $u$ and $v$ are approximately correct.

I found it exceptionally useful to visualize $u$ and $\partial
u/\partial t$ simultaneously via appropriate color mappings, with the
latter greatly amplified as in figure~\ref{fig:pi-deriv}. An
interactive simulation tool was essential for the discovery of stable
moving patterns ``in the wild''. Most work was performed on an 8-core
Xeon workstation; verification tests were also performed on a PowerPC
workstation. The simulation was implemented in C, divided into
anywhere from 2 to 16 execution threads via pthreads. The data set was
partitioned in one dimension only (stripes, each overlapping by two
rows with the two neighbor stripes); all inter-thread data exchange
was via shared memory.

Additional insight was gained from 1-D and 3-D numerical simulations
(the latter from the website at \cite{lidbeck2007}).

\section{Principal Findings}

I noted all of the pattern types reported in Pearson
\cite{pearson1993}, including several parameter values at which two or
more of Pearson's types coexist; many examples are exhibited at
\cite{munafo2009gallery}. In addition I observed three new pattern
types, illustrated in figure~\ref{fig:types-nx} and named $\nu$, $\xi$
and $\pi$ to extend Pearson's classification letters.

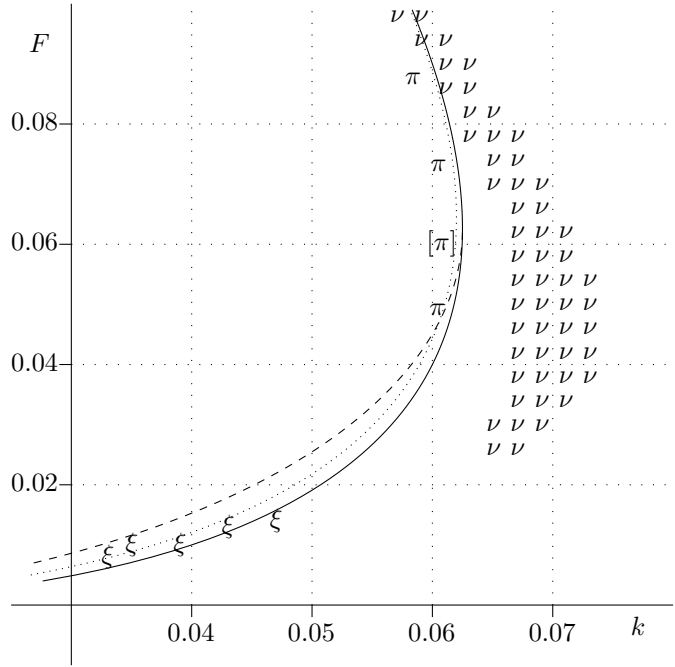
\begin{figure}
  \begin{center}
 \setlength{\unitlength}{0.8 cm}
    \begin{picture}(11.0, 11.0)(0,0)
    \begin{tikzpicture}[scale=0.8]

      \path[coordinate] (1.0,  0.0) coordinate(A)
                        (1.0, 11.0) coordinate(B)
                        ( 0.0, 1.0) coordinate(C)
                        (11.0, 1.0) coordinate(D);
      \draw[] (A) -- (B);
      \draw[] (C) -- (D);
      \draw[loosely dotted] (3.0, 1.0) -- (3.0, 11.0);
      \draw[] (3.0, 0.8) -- (3.0, 1.0);
         \put(2.6, 0.4){$0.04$};
      \draw[loosely dotted] (5.0, 1.0) -- (5.0, 11.0);
      \draw[] (5.0, 0.8) -- (5.0, 1.0);
         \put(4.6, 0.4){$0.05$};
      \draw[loosely dotted] (7.0, 1.0) -- (7.0, 11.0);
      \draw[] (7.0, 0.8) -- (7.0, 1.0);
         \put(6.6, 0.4){$0.06$};
      \draw[loosely dotted] (9.0, 1.0) -- (9.0, 11.0);
      \draw[] (9.0, 0.8) -- (9.0, 1.0);
         \put(8.6, 0.4){$0.07$};
      \draw[loosely dotted] (1.0, 3.0) -- (11.0, 3.0);
      \draw[] (0.8, 3.0) -- (1.0, 3.0);
         \put(0.0, 2.9){$0.02$};
      \draw[loosely dotted] (1.0, 5.0) -- (11.0, 5.0);
      \draw[] (0.8, 5.0) -- (1.0, 5.0);
         \put(0.0, 4.9){$0.04$};
      \draw[loosely dotted] (1.0, 7.0) -- (11.0, 7.0);
      \draw[] (0.8, 7.0) -- (1.0, 7.0);
         \put(0.0, 6.9){$0.06$};
      \draw[loosely dotted] (1.0, 9.0) -- (11.0, 9.0);
      \draw[] (0.8, 9.0) -- (1.0, 9.0);
         \put(0.0, 8.9){$0.08$};
      \put(10.3, 0.5){$k$};
      \put(0.3, 10.2){$F$};
\draw[] (0.524555, 1.400000)
       -- (1.071068, 1.500000)
       -- (1.545967, 1.600000)
       -- (1.966600, 1.700000)
       -- (2.344272, 1.800000)
       -- (2.686833, 1.900000)
       -- (3.000000, 2.000000)
       -- (3.288088, 2.100000)
       -- (3.554451, 2.200000)
       -- (3.801754, 2.300000)
       -- (4.032160, 2.400000)
       -- (4.247449, 2.500000)
       -- (4.449111, 2.600000)
       -- (4.638405, 2.700000)
       -- (4.816408, 2.800000)
       -- (4.984049, 2.900000)
       -- (5.142136, 3.000000)
       -- (5.291377, 3.100000)
       -- (5.432397, 3.200000)
       -- (5.565751, 3.300000)
       -- (5.691933, 3.400000)
       -- (5.811388, 3.500000)
       -- (5.924515, 3.600000)
       -- (6.031677, 3.700000)
       -- (6.133201, 3.800000)
       -- (6.229386, 3.900000)
       -- (6.320508, 4.000000)
       -- (6.406817, 4.100000)
       -- (6.488544, 4.200000)
       -- (6.565902, 4.300000)
       -- (6.639089, 4.400000)
       -- (6.708287, 4.500000)
       -- (6.773666, 4.600000)
       -- (6.835384, 4.700000)
       -- (6.893589, 4.800000)
       -- (6.948418, 4.900000)
       -- (7.000000, 5.000000)
       -- (7.048457, 5.100000)
       -- (7.093902, 5.200000)
       -- (7.136441, 5.300000)
       -- (7.176177, 5.400000)
       -- (7.213203, 5.500000)
       -- (7.247611, 5.600000)
       -- (7.279483, 5.700000)
       -- (7.308902, 5.800000)
       -- (7.335944, 5.900000)
       -- (7.360680, 6.000000)
       -- (7.383180, 6.100000)
       -- (7.403509, 6.200000)
       -- (7.421729, 6.300000)
       -- (7.437900, 6.400000)
       -- (7.452079, 6.500000)
       -- (7.464319, 6.600000)
       -- (7.474673, 6.700000)
       -- (7.483189, 6.800000)
       -- (7.489916, 6.900000)
       -- (7.494897, 7.000000)
       -- (7.498178, 7.100000)
       -- (7.499799, 7.200000)
       -- (7.499801, 7.300000)
       -- (7.498221, 7.400000)
       -- (7.495098, 7.500000)
       -- (7.490465, 7.600000)
       -- (7.484358, 7.700000)
       -- (7.476810, 7.800000)
       -- (7.467851, 7.900000)
       -- (7.457513, 8.000000)
       -- (7.445825, 8.100000)
       -- (7.432816, 8.200000)
       -- (7.418512, 8.300000)
       -- (7.402941, 8.400000)
       -- (7.386128, 8.500000)
       -- (7.368098, 8.600000)
       -- (7.348874, 8.700000)
       -- (7.328480, 8.800000)
       -- (7.306939, 8.900000)
       -- (7.284271, 9.000000)
       -- (7.260499, 9.100000)
       -- (7.235642, 9.200000)
       -- (7.209721, 9.300000)
       -- (7.182753, 9.400000)
       -- (7.154759, 9.500000)
       -- (7.125757, 9.600000)
       -- (7.095762, 9.700000)
       -- (7.064794, 9.800000)
       -- (7.032868, 9.900000)
       -- (7.000000, 10.000000)
       -- (6.966206, 10.100000)
       -- (6.931502, 10.200000)
       -- (6.895901, 10.300000)
       -- (6.859419, 10.400000)
       -- (6.822070, 10.500000)
       -- (6.783867, 10.600000)
       -- (6.744823, 10.700000)
       -- (6.704952, 10.800000)
       -- (6.664265, 10.900000);
\draw[dashed] (0.375156, 1.700000)
       -- (0.773589, 1.800000)
       -- (1.143118, 1.900000)
       -- (1.487850, 2.000000)
       -- (1.810956, 2.100000)
       -- (2.114936, 2.200000)
       -- (2.401812, 2.300000)
       -- (2.673245, 2.400000)
       -- (2.930624, 2.500000)
       -- (3.175121, 2.600000)
       -- (3.407739, 2.700000)
       -- (3.629344, 2.800000)
       -- (3.840694, 2.900000)
       -- (4.042442, 3.000000)
       -- (4.235181, 3.100000)
       -- (4.419423, 3.200000)
       -- (4.595635, 3.300000)
       -- (4.764234, 3.400000)
       -- (4.925595, 3.500000)
       -- (5.080054, 3.600000)
       -- (5.227922, 3.700000)
       -- (5.369479, 3.800000)
       -- (5.504983, 3.900000)
       -- (5.634667, 4.000000)
       -- (5.758747, 4.100000)
       -- (5.877426, 4.200000)
       -- (5.990882, 4.300000)
       -- (6.099287, 4.400000)
       -- (6.202797, 4.500000)
       -- (6.301561, 4.600000)
       -- (6.395709, 4.700000)
       -- (6.485368, 4.800000)
       -- (6.570653, 4.900000)
       -- (6.651676, 5.000000)
       -- (6.728535, 5.100000)
       -- (6.801324, 5.200000)
       -- (6.870130, 5.300000)
       -- (6.935035, 5.400000)
       -- (6.996113, 5.500000)
       -- (7.053433, 5.600000)
       -- (7.107066, 5.700000)
       -- (7.157064, 5.800000)
       -- (7.203489, 5.900000)
       -- (7.246387, 6.000000)
       -- (7.285811, 6.100000)
       -- (7.321797, 6.200000)
       -- (7.354391, 6.300000)
       -- (7.383623, 6.400000)
       -- (7.409529, 6.500000)
       -- (7.432134, 6.600000)
       -- (7.451463, 6.700000)
       -- (7.467537, 6.800000)
       -- (7.480377, 6.900000)
       -- (7.489995, 7.000000)
       -- (7.496399, 7.100000)
       -- (7.499601, 7.200000);
\draw[dotted] (0.326593, 1.500000)
       -- (0.825715, 1.600000)
       -- (1.271406, 1.700000)
       -- (1.674129, 1.800000)
       -- (2.041291, 1.900000)
       -- (2.378371, 2.000000)
       -- (2.689567, 2.100000)
       -- (2.978173, 2.200000)
       -- (3.246841, 2.300000)
       -- (3.497735, 2.400000)
       -- (3.732659, 2.500000)
       -- (3.953133, 2.600000)
       -- (4.160444, 2.700000)
       -- (4.355704, 2.800000)
       -- (4.539876, 2.900000)
       -- (4.713806, 3.000000)
       -- (4.878230, 3.100000)
       -- (5.033809, 3.200000)
       -- (5.181130, 3.300000)
       -- (5.320712, 3.400000)
       -- (5.453030, 3.500000)
       -- (5.578507, 3.600000)
       -- (5.697535, 3.700000)
       -- (5.810462, 3.800000)
       -- (5.917612, 3.900000)
       -- (6.019275, 4.000000)
       -- (6.115727, 4.100000)
       -- (6.207216, 4.200000)
       -- (6.293973, 4.300000)
       -- (6.376210, 4.400000)
       -- (6.454124, 4.500000)
       -- (6.527904, 4.600000)
       -- (6.597719, 4.700000)
       -- (6.663731, 4.800000)
       -- (6.726088, 4.900000)
       -- (6.784934, 5.000000)
       -- (6.840400, 5.100000)
       -- (6.892607, 5.200000)
       -- (6.941676, 5.300000)
       -- (6.987717, 5.400000)
       -- (7.030835, 5.500000)
       -- (7.071124, 5.600000)
       -- (7.108679, 5.700000)
       -- (7.143590, 5.800000)
       -- (7.175938, 5.900000)
       -- (7.205805, 6.000000)
       -- (7.233266, 6.100000)
       -- (7.258395, 6.200000)
       -- (7.281254, 6.300000)
       -- (7.301914, 6.400000)
       -- (7.320436, 6.500000)
       -- (7.336878, 6.600000)
       -- (7.351296, 6.700000)
       -- (7.363748, 6.800000)
       -- (7.374284, 6.900000)
       -- (7.382952, 7.000000)
       -- (7.389803, 7.100000)
       -- (7.394881, 7.200000)
       -- (7.398229, 7.300000)
       -- (7.399892, 7.400000)
       -- (7.399907, 7.500000)
       -- (7.398315, 7.600000)
       -- (7.395154, 7.700000)
       -- (7.390458, 7.800000)
       -- (7.384266, 7.900000)
       -- (7.376606, 8.000000)
       -- (7.367513, 8.100000)
       -- (7.357018, 8.200000)
       -- (7.345152, 8.300000)
       -- (7.331942, 8.400000)
       -- (7.317418, 8.500000)
       -- (7.301605, 8.600000)
       -- (7.284531, 8.700000)
       -- (7.266221, 8.800000)
       -- (7.246699, 8.900000)
       -- (7.225990, 9.000000)
       -- (7.204115, 9.100000)
       -- (7.181097, 9.200000)
       -- (7.156959, 9.300000)
       -- (7.131719, 9.400000)
       -- (7.105401, 9.500000)
       -- (7.078023, 9.600000)
       -- (7.049604, 9.700000)
       -- (7.020160, 9.800000)
       -- (6.989714, 9.900000)
       -- (6.958280, 10.000000)
       -- (6.925876, 10.100000)
       -- (6.892519, 10.200000)
       -- (6.858224, 10.300000)
       -- (6.823010, 10.400000)
       -- (6.786888, 10.500000)
       -- (6.749874, 10.600000)
       -- (6.711983, 10.700000)
       -- (6.673228, 10.800000)
       -- (6.633624, 10.900000);
  \put(6.3, 10.7){$\nu$};
  \put(6.7, 10.7){$\nu$};
  \put(6.7, 10.3){$\nu$};
  \put(7.1, 10.3){$\nu$};
  \put(7.1, 9.9){$\nu$};
  \put(7.5, 9.9){$\nu$};
  \put(7.1, 9.5){$\nu$};
  \put(7.5, 9.5){$\nu$};
  \put(7.5, 9.1){$\nu$};
  \put(7.9, 9.1){$\nu$};
  \put(7.5, 8.7){$\nu$};
  \put(7.9, 8.7){$\nu$};
  \put(8.3, 8.7){$\nu$};
  \put(7.9, 8.3){$\nu$};
  \put(8.3, 8.3){$\nu$};
  \put(7.9, 7.9){$\nu$};
  \put(8.3, 7.9){$\nu$};
  \put(8.7, 7.9){$\nu$};
  \put(8.3, 7.5){$\nu$};
  \put(8.7, 7.5){$\nu$};
  \put(8.3, 7.1){$\nu$};
  \put(8.7, 7.1){$\nu$};
  \put(9.1, 7.1){$\nu$};
  \put(8.3, 6.7){$\nu$};
  \put(8.7, 6.7){$\nu$};
  \put(9.1, 6.7){$\nu$};
  \put(8.3, 6.3){$\nu$};
  \put(8.7, 6.3){$\nu$};
  \put(9.1, 6.3){$\nu$};
  \put(9.5, 6.3){$\nu$};
  \put(8.3, 5.9){$\nu$};
  \put(8.7, 5.9){$\nu$};
  \put(9.1, 5.9){$\nu$};
  \put(9.5, 5.9){$\nu$};
  \put(8.3, 5.5){$\nu$};
  \put(8.7, 5.5){$\nu$};
  \put(9.1, 5.5){$\nu$};
  \put(9.5, 5.5){$\nu$};
  \put(8.3, 5.1){$\nu$};
  \put(8.7, 5.1){$\nu$};
  \put(9.1, 5.1){$\nu$};
  \put(9.5, 5.1){$\nu$};
  \put(8.3, 4.7){$\nu$};
  \put(8.7, 4.7){$\nu$};
  \put(9.1, 4.7){$\nu$};
  \put(9.5, 4.7){$\nu$};
  \put(8.3, 4.3){$\nu$};
  \put(8.7, 4.3){$\nu$};
  \put(9.1, 4.3){$\nu$};
  \put(7.9, 3.9){$\nu$};
  \put(8.3, 3.9){$\nu$};
  \put(8.7, 3.9){$\nu$};
  \put(7.9, 3.5){$\nu$};
  \put(8.3, 3.5){$\nu$};
  \put(6.548, 9.66){$\pi$};
  \put(6.966, 8.22){$\pi$};
  \put(6.9, 6.9){$[\pi]$};
  \put(6.952, 5.82){$\pi$};
  \put(4.3, 2.3){$\xi$};
  \put(3.5, 2.2){$\xi$};
  \put(1.9, 1.9){$\xi$};
  \put(2.7, 1.9){$\xi$};
  \put(1.5, 1.7){$\xi$};

    \end{tikzpicture}
    \end{picture}
 \setlength{\unitlength}{1.0cm}
    \caption{
      Parameter values that produce pattern types $\nu$, $\xi$ and $\pi$.
      The solid curve is the saddle-node bifurcation,
      the dashed curve is the Hopf bifurcation,
      and the dotted curve is the Turing instability threshold.
      The area around $(F=0.06, k=0.0609)$ is indicated by $[\pi]$.
      \label{fig:param-space}
    }
  \end{center}
\end{figure}

\begin{figure}
  \centering
    \begin{tabular}[]{c c}

    \begin{subfigure}[b]{0.2\textwidth}
      \includegraphics[width=\textwidth]{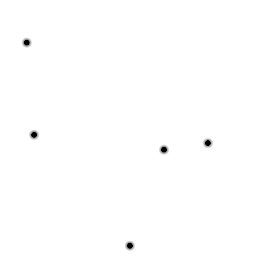}
      \caption{$F=0.040$, $k=0.070$}
      \label{fig:type-nu}
    \end{subfigure}
    &
    \begin{subfigure}[b]{0.2\textwidth}
      \includegraphics[width=\textwidth]{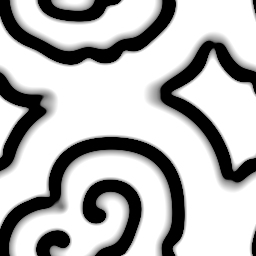}
      \caption{$F=0.014$, $k=0.047$}
      \label{fig:type-xi}
    \end{subfigure}
    \\
    \begin{subfigure}[b]{0.2\textwidth}
      \includegraphics[width=\textwidth]{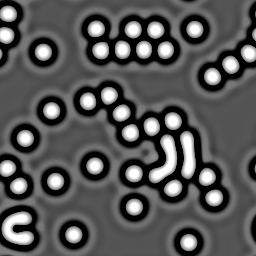}
      \caption{$F=0.06$, $k=0.0609$}
      \label{fig:wild-pi}
    \end{subfigure}
    &
    \begin{subfigure}[b]{0.2\textwidth}
    \begin{tikzpicture}[scale=1.0]
      \draw (0,0) rectangle (3.6,2.5);

      \filldraw[fill=white, draw=black] (0.1,1.7) rectangle (0.8,2.4);
      \put(1.0,1.9){$u < 0.40$};

      \filldraw[fill=white!70!black, draw=white!70!black]
                                        (0.1,0.9) rectangle (0.8,1.6);

      \path[coordinate] (0.1,1.1) coordinate(A)
                        (0.1,0.9) coordinate(B)
                        (0.3,0.9) coordinate(C)
                        (0.8,1.4) coordinate(D)
                        (0.8,1.6) coordinate(E)
                        (0.6,1.6) coordinate(F);
      \filldraw[fill=white!50!black, draw=white!50!black]
                     (A) -- (B) -- (C) -- (D) -- (E) -- (F) -- cycle;

      \path[coordinate] (0.8,0.9) coordinate(H);
      \filldraw[fill=white!30!black, draw=white!30!black]
                                (C) -- (H) -- (D) -- cycle;
      \put(1.0,1.1){$0.40 < u < 0.44$};

      \filldraw[fill=black]             (0.1,0.1) rectangle (0.8,0.8);
      \put(1.0,0.3){$0.44 < u$};
    \end{tikzpicture}
    \caption{shading key}
    \end{subfigure}
    \end{tabular}
    \caption{
      Pattern types $\nu$ (inert autosolitons),
      $\xi$ (B-Z targets and spirals), and
      $\pi$ (stable, stationary and moving localized patterns).
      This color key applies to all subsequent figures.
      \label{fig:types-nx}
    }
\end{figure}
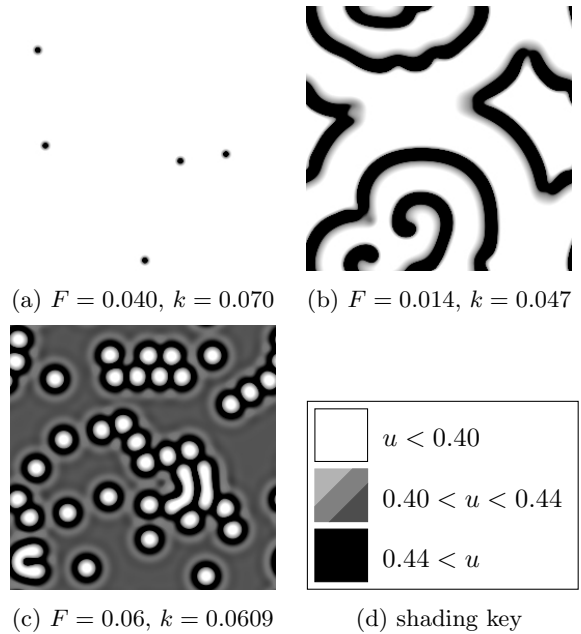

Type $\nu$ is found throughout a large part of the area labeled \typeR/
in \cite{pearson1993}, bordering on the regions $\varepsilon$,
$\lambda$, and $\mu$. Here we have stable spots (called solitons or
autosolitons by others \cite{muratov2000, stollenwerk2008}) that do
not multiply. These spots are static only in isolation. Pairs and
groups drift apart from one another, at a rate that diminishes sharply
with distance. The velocity of two spots drifting apart in a
sufficiently large and otherwise empty domain is modeled fairly well
by an exponentially decaying rate: velocity $c \approx K
e^{-rd}~\dlu/~\dtu/^{-1}$ where $d$ is inter-spot distance, $K$ and $r$
constant. For $(F = 0.04, k = 0.07)$ as shown here, $r \approx
46~\dlu/^{-1}$: the spots drift about half as fast with every added
$0.015~\dlu/$ of inter-spot distance. $r$ increases with $k$ and with
$F$; for example, at $(F = 0.04, k = 0.072)$, $r \approx 47$; and at
$(F = 0.08, k = 0.066)$, $r \approx 67$. Given enough time and a small
domain size, motion ceases with spots roughly equidistant; such
multi-spot patterns are the subject of \cite{wei2003}. At lower $F$
values, particularly for $F < 0.04$, spot-like starting patterns
produce spots that oscillate with a characteristic frequency varying
slightly with $k$ and $F$, and a damping rate that diminishes sharply
as $(F, k)$ approaches the right edge of the type $\nu$ region in
figure~\ref{fig:param-space}. Near this frontier the stable spot
becomes smaller in diameter and the central peak has $(u, v)$ farther
from $(1, 0)$. Also in figure~\ref{fig:param-space} we see that at
higher $F$ values type $\nu$ occurs on both sides of the saddle-node
bifurcation line. As Mazin et. al. found in one dimension
\cite{mazin1996}, at higher $F$ values the red state overpowers the
blue state in a substantial band to the left of the saddle-node line;
however it is not strong enough to extinguish small spots. For this
reason, parameters in this band support solitons that do not grow to
fill the space with the blue state.

Type $\xi$ is found in a modest-sized area of the parameter space
which includes $(F=0.014, k=0.047)$ and $(F=0.008, k=0.033)$. As shown
in figure~\ref{fig:param-space}, type $\xi$ appears on both sides of
the saddle-node bifurcation. It also appears on both sides of the
subcritical/supercritical boundary at $k = 9/256 \approx 0.035$ (when
$k<9/256$, the model without diffusion oscillates indefinitely; see
\cite{gray1994} and \cite{mazin1996}). Type $\xi$ patterns very
closely resemble the B-Z (Belousov-Zhabotinsky) reaction in a Petri
dish. An initial pattern containing a traveling wave front with free
ends produces two spiral seeds and usually results in sustained
activity. At some parameter values the waves develop flaws similar to
the segmented waves shown in \cite{yang2005} (which is discussing a
CDIMA system); these flaws contribute to the production of more spiral
seeds. Occasionally, double spirals also occur. A self-sustaining
population of spiral seeds is needed to maintain the pattern, and the
density of seeds varies with $k$ and $F$. The longevity of the pattern
is highly dependent on the size of the domain. At $(F=0.014,
k=0.047)$, twenty starting patterns like that in
figure~\ref{fig:uskate-generation} (a $1.79~\dlu/ \times 1.79~\dlu/$
domain) had an average life-span of 5400 \dtu/ before all waves
died out. Twenty trials in a $6.7~\dlu/ \times 6.7~\dlu/$ domain with
starting patterns of the same density all lasted longer than $10^6$
\dtu/.

Type $\pi$ is the most novel and the subject of the rest of this
paper. I found patterns in this category for $F$ values ranging from
$0.04$ to $0.09$, in a very thin band of $k$ values;
figure~\ref{fig:param-space} shows four representative locations in this
band. The band runs roughly parallel to the saddle-node and Turing
lines, and to the left (lower $k$ values) of both. It is also at lower
$k$ values than the area investigated by McGough and Riley
\cite{mcgough2004}.

In \cite{pearson1993} Pearson explored some parameter values near this
area, initializing the system with red state $(u=1, v=0)$ and a small
central rectangle of $(u=1/2, v=1/4)$, and found that the system
evolved to a homogeneous blue state; these were designated \typeB/.
Mazin et. al. \cite{mazin1996} using a similar starting pattern in
one-dimensional simulations, explored a greater number of $F$ and $k$
combinations, and found many for which stable localized structures
result; these were designated \typeL/.

At all $F$ values the $\pi$ band is very narrow in $k$ as compared to
its distance from the bifurcation lines. For example, when $F=0.06$,
the saddle-node bifurcation is at $k\approx0.06247$, the Hopf
bifurcation is at $k\approx0.06245$, the Turing bifurcation is at
$k\approx0.06191$, and the range of valid $k$ values for the object in
the lower-left of figure~\ref{fig:pi-valu} is $0.06087\pm0.00001 \leq
k \leq 0.06098\pm0.00001$. Thus the width of the $\pi$ band near
$F=0.06$ is about $1.1\times10^{-4}$ in units of $k$, while the
distance to the nearby Turing region is ten times as great.

Pearson's pattern type $\iota$ (see figure 2 in \cite{pearson1993})
shares some qualities with type $\pi$. It is in region \typeL/ of Mazin
et. al. \cite{mazin1996}, but its lack of solitary spots indicates it
is probably slightly outside the $\pi$ band.

\section{Stability and motion of type \texorpdfstring{$\pi$}{pi} Patterns}

\label{sec-stability}

\begin{figure}
  \centering
  \begin{subfigure}[b]{0.2\textwidth}
    \includegraphics[width=\textwidth]{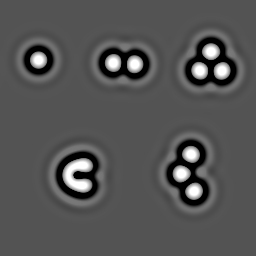}
    \caption{$u$}
    \label{fig:pi-valu}
  \end{subfigure}
  \hspace{2mm}
  \begin{subfigure}[b]{0.2\textwidth}
    \includegraphics[width=\textwidth]{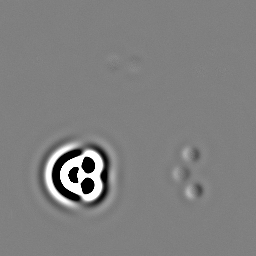}
    \caption{$\partial u/\partial t$}
    \label{fig:pi-deriv}
  \end{subfigure}
  \hspace{2mm}
  \begin{subfigure}[b]{0.2\textwidth}
    \includegraphics[width=\textwidth]{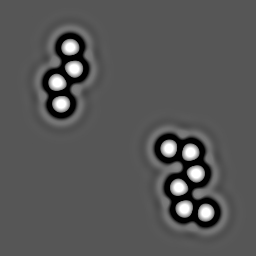}
    \caption{rotating patterns}
    \label{fig:rotate}
  \end{subfigure}
  \hspace{2mm}
  \begin{subfigure}[b]{0.2\textwidth}
    \includegraphics[width=\textwidth]{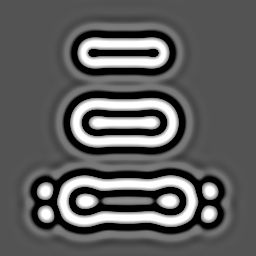}
    \caption{stripelike patterns}
    \label{fig:stripes}
  \end{subfigure}
  \caption{
    (a),(b): Comparison of $u$ to its time derivative for pattern type $\pi$.
    Coloring for all but (b) is the same as above.
    In image (b), white represents
    $\partial u/\partial t > 2\times10^{-6}$,
    shades of gray where $|\partial u/\partial t| < 2\times10^{-6}$,
    and black otherwise.
    (c): Two stable rotating patterns.
    (d): Three linear growing patterns.
    $(F = 0.06, k=0.0609)$ for all figures;
    details in section~\ref{sec-stability}.
    \label{fig:partials}
  }
\end{figure}

Figure~\ref{fig:pi-valu} shows five patterns at $F = 0.06$ and $k =
0.0609$. The three in the upper row are stable non-moving patterns.
Contrast has been exaggerated in the area of $0.40 < u < 0.44$ to show
the concentric rings or ``halos'' that surround all patterns in
systems with these parameters. The halos are concentric stationary
waves of alternating sign superimposed on the homogeneous state values
$(u_{h3},v_{h3})$ in \eqref{uv-h3}. The full range of $u$ and $v$
levels for type $\pi$ patterns at $F=0.06, k=0.0609$ is $0.29 < u < 0.86$
and $0.01 < v < 0.43$.

The other two patterns in figure \ref{fig:pi-valu} move to the left,
indefinitely at constant speed; the three-spot pattern moves at about
1 \dlu/ per $8.5 \times 10^6~\dtu/$, and the \textsf{U}-shaped
pattern at about 1 \dlu/ per $6.2 \times 10^4~\dtu/$.

All of the patterns in \ref{fig:pi-valu} arise frequently from random
starting patterns, and are resilient to noise and other perturbations.
If any of them is perturbed by shifting half of the pattern in any
direction a distance on the order of 0.02 \dlu/, further
simulation results in a return to the canonical forms shown here.

Figure~\ref{fig:pi-deriv} shows the time derivative of $u$ for the
same five patterns. The motion of the two patterns in the bottom half
of the figure is clear.

Figure~\ref{fig:rotate} shows two rotating patterns. Both rotate
clockwise; the four-spot pattern performs one full revolution in about
$1.6 \times 10^7~\dtu/$; the other takes about $1.2 \times 10^7~\dtu/$.
Disruptions to these will cause a momentary change in rotational
velocity followed by a return to the normal rotation rate after the
spots return to the stable alignment.

Many stripelike patterns like those in figure~\ref{fig:stripes} arise
from random starting patterns; one-ended forms are more common than
the two-ended versions shown here. They are stable in the central
linear section, in the direction perpendicular to their length, but
are unstable in the other dimension. Any deviation from a straight
line will increase, first slowly and then with increasing speed,
forming meanders like those of a river in a floodplain. These linear
forms grow at both ends. The top one is the fastest, each end grows at
about 1 \dlu/ per $6.7 \times 10^4~\dtu/$, slightly slower than the
speed of the \textsf{U}-shaped pattern in figure~\ref{fig:pi-valu}.
The second example grows at 1 \dlu/ per $1.45 \times 10^6~\dtu/$ from
each end. The third example has two spots at each end that make it
grow (at about 1 \dlu/ per $2 \times 10^6~\dtu/$); without these spots
the pattern shrinks at each end. When encountering other objects,
these stripelike patterns will change direction, stop growing, or
(frequently in the case of the topmost example) break down into some
other form, such as separate parallel stripes.

Systems that contain active growing stripes will usually grow to fill
all available space with spots and stripes, but the time taken and the
final proportion of spots to stripes is highly variable, and depends
on tiny details of the initial configuration. This is substantially
different from the behavior of stripes at other parameter values in
the Gray-Scott system, whose final density can generally be predicted
from the parameter values alone.

The central portions of the stripelike patterns in
figure~\ref{fig:stripes} suggest a connection with certain
one-di\-men\-sion\-al localized structures found by Mazin et. al in
\cite{mazin1996} at parameter values in their region \typeL/. In my
tests I found that the dimensions and levels of $u$ and $v$ of
features in 1-D simulations were effectively the same as
cross-sections of stripe-like patterns in 2-dimensional simulations
when the same parameter values are used.

\section{Variations with small changes to \texorpdfstring{$k$}{k}}

\label{sec-vary-k}

\begin{figure}
  \centering
  \begin{subfigure}[b]{0.2\textwidth}
    \includegraphics[width=\textwidth]{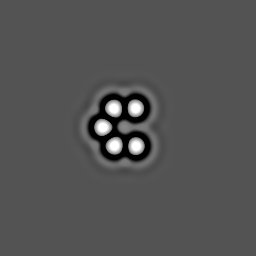}
    \caption{$k = 0.0609$}
    \label{fig:hex-k6090}
  \end{subfigure}
  \hspace{2mm}
  \begin{subfigure}[b]{0.2\textwidth}
    \includegraphics[width=\textwidth]{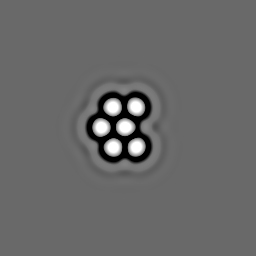}
    \caption{$k = 0.0611$}
    \label{fig:hex-k6110}
  \end{subfigure}
  \hspace{2mm}
  \begin{subfigure}[b]{0.2\textwidth}
    \includegraphics[width=\textwidth]{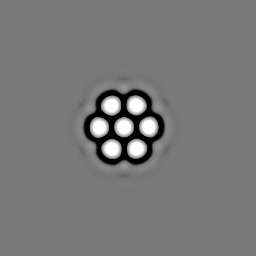}
    \caption{$k = 0.06135$}
    \label{fig:hex-k6135}
  \end{subfigure}

  \centering
  \begin{subfigure}[b]{0.2\textwidth}
    \includegraphics[width=\textwidth]{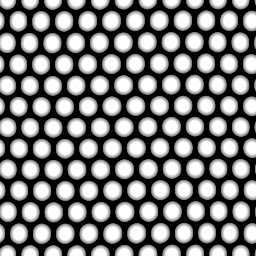}
    \caption{$k = 0.0615$}
    \label{fig:hex-k6150}
  \end{subfigure}
  \hspace{2mm}
  \begin{subfigure}[b]{0.2\textwidth}
    \includegraphics[width=\textwidth]{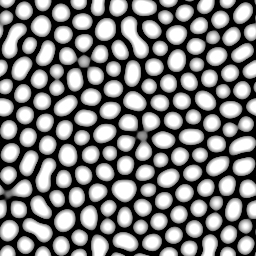}
    \caption{$k = 0.0620$}
    \label{fig:turing-k6200}
  \end{subfigure}
  \caption{
    (a) through (d) show the evolution of a starting pattern as $k$
    is raised progressively to three nearby values.
    For comparison, (e) shows a Turing pattern arising from
    low-level random noise.
    $F = 0.06$ for all figures;
    details in section~\ref{sec-vary-k}.
    \label{fig:hexes}
  }
\end{figure}

The patterns in figure~\ref{fig:pi-valu} are shown as they appear when
$F=0.06$ and $k=0.0609$. If $k$ is diminished below $0.06087$, the
\textsf{U}-shaped moving pattern is no longer viable and decays into a
single spot. When $k < 0.06062$ the single spot shown in the
upper-left of figure~\ref{fig:pi-valu} is no longer stable and quickly
evolves to the homogeneous state, however the other three patterns
made up of spots continue to exist. When $k$ is lowered below
$0.06060$, the patterns in the top center and lower right vanish, but
the triangular three-spot pattern remains. Below $0.06057$ the
triangular pattern vanishes, but a 7-spot arrangement similar to that
in figure~\ref{fig:hex-k6135} is still viable. The 7-spot pattern dies
out when $k < 0.06055$.

At $F=0.06, k=0.06090$ the 5-spot pattern in figure~\ref{fig:hex-k6090}
is stable and moves to the left slowly. Using this as the starting
pattern, $k$ is increased to 0.06110, and a spot forms in the center to
make the 6-spot pattern in figure~\ref{fig:hex-k6110}. This pattern is
also left-moving and stable. Increasing $k$ further to $0.06135$, a
seventh spot appears (figure~\ref{fig:hex-k6135}), and motion stops
due to the attainment of symmetry. Throughout this process the spots
shift away from one another slightly to attain a new steady state,
with a slightly greater inter-spot spacing with each increase in $k$.

Increasing $k$ to $0.06150$, new spots appear on all sides, quickly
growing to produce the uniform hexagonal grid in
figure~\ref{fig:hex-k6150}.

Groups of spots are no longer stable at these relatively high $k$
values -- if $k$ is increased more gradually from 0.06135 to 0.06150
(for example in 15 equal steps at intervals of $5000~\dtu/$), some or
all of the spots will have time to swell into elongated ``stripes''
of high $u$. Even at $k = 0.06135$ the seven-spot pattern shown in
\ref{fig:hex-k6135} is barely stable, and groupings of more than 7
spots produce stripelike patterns.

Considered together, the range of different behaviors described thus
far represents a very wide spectrum of phenomena in the
relatively narrow range $0.06055 < k < 0.06150$. A similar spectrum
is found at higher and lower $F$ values, although varying in details.
For example, the behavior shown in figure~\ref{fig:hex-k6150} has been
observed at several values of $F$ from 0.046 to 0.0652; at higher $F$
values a greater change in $k$ is needed to precipitate the growth of
the hexagonal pattern, and it appears that at significantly higher $F$
the phenomenon cannot be produced at all. The \textsf{U}-shaped moving
pattern has been found to be stable at $F$ values from 0.0492 to
0.0876, with $k$ varying as illustrated in figure~\ref{fig:param-space}.

At $F=0.06, k=0.062$, the Turing effect is pres\-ent. Here, no initial
spots are needed to generate a pattern of spots and/or stripes. As
defined by \cite{turing1952} (see also the introductory section of
\cite{mazin1996}) Turing patterns arise spontaneously from random
noise of arbitrarily low initial amplitude, diffusion plays an active
role in the destabilization of the initial state, and a characteristic
wavelength (spot size and/or stripe width) exists that is independent
of the system size. Figure~\ref{fig:turing-k6200} shows a typical
Turing pattern in the 2-D Gray-Scott system; it was produced from a
starting pattern of ``white'' noise with amplitude $10^{-4}$
superimposed on the blue state $(u_{h3},v_{h3})$ from \eqref{uv-h3}.
We see similarities, such as hexagonal arrangement of spots and a
similar spot size and spacing, to figure~\ref{fig:hex-k6150}. However,
many grain boundaries, short stripes and other differences in detail
are present. This is typical of patterns throughout the Turing domain,
which at $F=0.06$ includes $k$ values in the range $0.06191<k<0.06245$
(see \cite{mazin1996} for a derivation of the formulas for these
values; $k=0.06245$ is the Hopf bifurcation).

\section{Distance interactions and more exotic patterns}

\label{sec-complex}

\begin{figure}
  \centering
  \begin{subfigure}[b]{0.2\textwidth}
    \includegraphics[width=\textwidth]{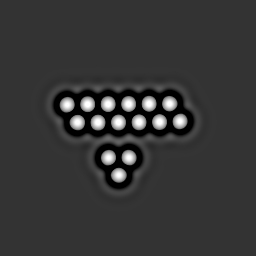}
    \caption{$k = 0.06058$}
    \label{fig:expan-6058}
  \end{subfigure}
  \hspace{2mm}
  \begin{subfigure}[b]{0.2\textwidth}
    \includegraphics[width=\textwidth]{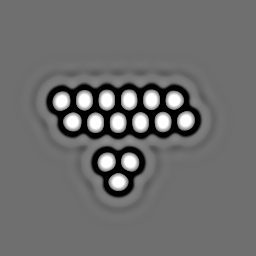}
    \caption{$k = 0.06115$}
    \label{fig:expan-6115}
  \end{subfigure}
  \hspace{2mm}
  \begin{subfigure}[b]{0.2\textwidth}
    \includegraphics[width=\textwidth]{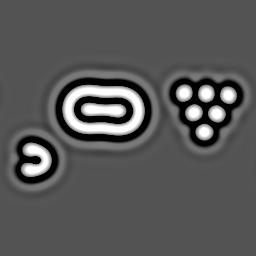}
    \caption{Interaction (before)}
    \label{fig:int-before}
  \end{subfigure}
  \hspace{2mm}
  \begin{subfigure}[b]{0.2\textwidth}
    \includegraphics[width=\textwidth]{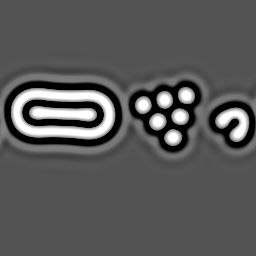}
    \caption{Interaction (after)}
    \label{fig:int-after}
  \end{subfigure}
  \hspace{2mm}
  \caption{
    (a),(b): Lattice spacing variation;
    (c),(d): An interaction between three patterns.
    $(F = 0.06, k=0.0609)$ except as otherwise noted;
    details in section~\ref{sec-complex}.
    \label{fig:spacing}
  }
\end{figure}

Figures~\ref{fig:expan-6058} and \ref{fig:expan-6115} show two frames
of a simulation in which $k$ was gradually increased from 0.06058 to
0.06110 over $10^6~\dtu/$. The overall length of the horizontal
row of spots increases with $k$, and the smaller 3-spot pattern moves
slightly away from it. The entire set of 15 spots also rotates very
slowly as a unit. Similar interactions between patterns that are not
in direct contact are very common. Decreasing $k$ back to 0.06058 over
a similar time period causes the pattern to return to the state in
\ref{fig:expan-6058}.

In figure~\ref{fig:int-before}, the \textsf{U}-shaped feature is moving to the
right and slightly downward. It interacts weakly with the stripelike
feature, then more strongly with the triangular array of spots to the
right, and emerges traveling on a diagonal path towards the
upper-right. An animation of this is available at \cite{munafo2009web}.

\begin{figure}
  \centering
    \begin{tabular}[]{c c c}

      \begin{subfigure}[b]{0.11\textwidth}
        \includegraphics[width=\textwidth]{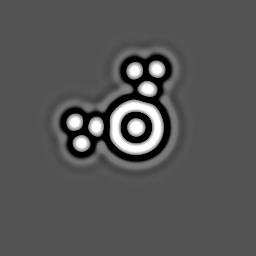}
        \caption{}
        \label{fig:target-2x3-lop}
      \end{subfigure}
      &
      \begin{subfigure}[b]{0.11\textwidth}
        \includegraphics[width=\textwidth]{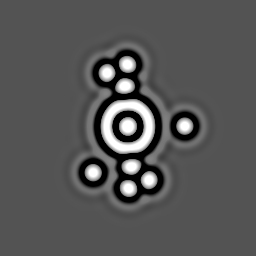}
        \caption{}
        \label{fig:targ-2x3-spots1}
      \end{subfigure}
      &
      \begin{subfigure}[b]{0.11\textwidth}
        \includegraphics[width=\textwidth]{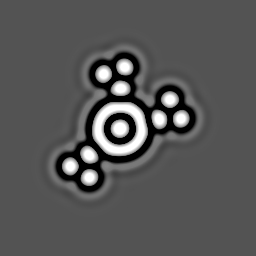}
        \caption{}
        \label{fig:target-3x3-lop}
      \end{subfigure}
      \\
      \begin{subfigure}[b]{0.11\textwidth}
        \includegraphics[width=\textwidth]{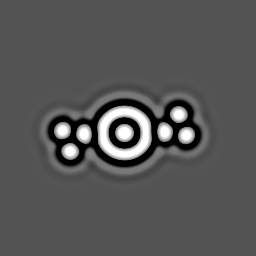}
        \caption{}
        \label{fig:targ-2x3}
      \end{subfigure}
      &
      \begin{subfigure}[b]{0.11\textwidth}
        \includegraphics[width=\textwidth]{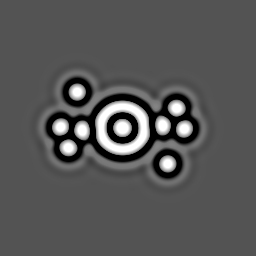}
        \caption{}
        \label{fig:targ-2x(3+1)}
      \end{subfigure}
      &
      \begin{subfigure}[b]{0.11\textwidth}
        \includegraphics[width=\textwidth]{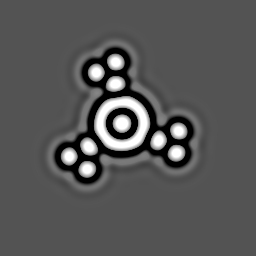}
        \caption{}
        \label{fig:target-3x3}
      \end{subfigure}
    \end{tabular}
    \begin{tabular}[]{c}
      \begin{subfigure}[b]{0.22\textwidth}
        \includegraphics[width=\textwidth]{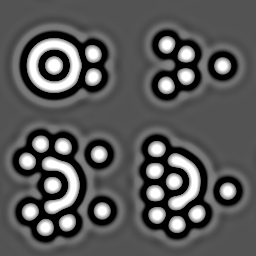}
        \caption{rare moving forms}
        \label{fig:four-spaceships}
      \end{subfigure}
    \end{tabular}
    \caption{
      (a)-(c) are precursors to (d)-(f), which are stable and rotate clockwise.
      Patterns in (g) move to the right; the asymmetrical pattern moves
      on a curved path. Details in section~\ref{sec-complex}.
      $(F = 0.06, k=0.0609)$ for all figures.
      \label{fig:exotic}
    }
\end{figure}

In figure~\ref{fig:exotic} are six rotating patterns with a ``target''
pattern in the center. The first three are asymmetrical; each of these
transforms into the corresponding symmetrical version; the center
shifts a bit in the process. All rotate clockwise. The solitary spots
in the second example are pushed by the triads. In
figure~\ref{fig:targ-2x3-spots1} one of the spots is stationary; after
becoming symmetrical as shown in \ref{fig:targ-2x(3+1)} both spots are
moving. Each of these solitary spots remains at the distance shown,
about twice the radius of the brightest surrounding halo, from the
annulus of the ``target'' and from the two closer spots of the triad
immediately following it.

Figure~\ref{fig:four-spaceships} shows several more moving patterns.
All move to the right. The pattern in the upper-left is an example of
many variations of the ``target'' with attached spots that arise from
random starting patterns. Most remain stable and move; symmetrical
forms move on a straight path and asymmetrical forms move on a curved
path. The ``target'' by itself is also stable and does not move. The
other three all involve spots that are being ``pushed'' ahead of a
larger pattern; in all cases the larger pattern moves faster if the
spots in front are removed. The asymmetrical pattern in the
lower-right moves on a curved path, bending to its right as it moves
forward. In the process the spot is pushed out of the way, and is left
behind after about the first quarter-turn of the moving pattern's
path. An animation of this is available at \cite{munafo2009web}.

\section{Similarity to Other Systems}

Several other authors describe spots with appearance and interaction
similar to the spots shown in this paper.

Schenk, et. al. \cite{schenk1998}, studying a two-component cubic
autocatalytic reaction-diffusion system, found parameter values for
which there is a stable spot solution with concentric rings of
alternating sign and diminishing magnitude. For certain parameter
values they found that the spots interact in a similar way to that
described in this paper; they showed some stable multi-spot
arrangements similar to my figure~\ref{fig:pi-valu} and proposed
several more. There are some differences, for example the pattern in
their figure 5(e) (a four-spot pattern in a \textsf{Y} configuration
with unequal angles) is not stable in the Gray-Scott system with
parameters $(F=0.06, k=0.0609)$, nor at several other locations I
tested throughout the $\pi$ region. I also found the pattern in their
figure 5(b) (three spots in a straight line) to be stable, whereas
they did not. Such differences might result from differences in the
spacing and relative amplitudes of the halos.

In work related to \cite{schenk1998}, a number of researchers
including Purwins et. al. \cite{purwins1999} have identified and
studied ``quasi-particles'' that can arrange themselves in a very
similar manner to my figures \ref{fig:pi-valu} and
\ref{fig:hex-k6150}. This can be seen clearly in figures 3 (d), (g),
and (h) of the web page article by Stollenwerk \cite{stollenwerk2008}.
Note in particular the halos around the spots in their figure 3(d).
These spots arose in laboratory AC gas plasma discharge experiments
that have other features reminiscent of patterns
seen in the two-dimensional Gray-Scott system.

Liehr, et. al. in \cite{liehr2000}, modeling a DC gas plasma discharge
experiment, began with a 3-component reaction-diffusion system, then
transformed the equations into a two-component form with a global
integration term. In numerical simulation (see their figures 3 and 5)
they achieved results similar to my figures \ref{fig:hex-k6135} and
\ref{fig:hex-k6150}. Note the halos and arrangement of the spots in
their patterns. Studying a similar 3-component system, Schenk in
\cite{schenk1999} shows several examples that include a
\emph{Zielscheiben-Struktur} (target structure) closely resembling the
target patterns in my figure~\ref{fig:exotic}. Spots with halos and
similar spot-to-spot interactions are also shown in that work. In both
of these 3-component systems, a single spot can have an intrinsic velocity
and display particle-like collision behavior in addition to the static
attraction and repulsion effects.

\section{Discussion}

The single spot in the upper-left of figure~\ref{fig:pi-valu} has a set of
concentric halos with progressively lower amplitude and alternating
sign. In multi-spot patterns, each spot tends to be found at a
location that coincides with the first positive-sign halos of
neighboring spots. In the growth of hexagonal arrays, the new spots
always appear at such locations.

In figure~\ref{fig:pi-valu}, the \textsf{U}-shaped moving pattern is
of similar size and shape to the three-spot pattern, and both tend to
return to the canonical dimensions shown here after distortion or
perturbation.

In more distant alignments such as that in figure~\ref{fig:expan-6058}
the spots are found at distances coinciding with each other's second
positive-signed surrounding halos. This is also evident in the
five-spot pattern in figure~\ref{fig:hex-k6090}, which remains as
shown when perturbed, rather than changing to a pentagonal ring or
some other arrangement.

In several instances (for example figure~\ref{fig:stripes}, all of
figure~\ref{fig:spacing}, and figure~\ref{fig:targ-2x3}) we see rows
of spots and other roughly linear features, halos that parallel these
features, and motion and interaction that follows the locations of
these roughly linear halos.

All of the foregoing suggest that areas of high $u$ and low $v$
produce a pattern of surrounding standing waves, and that these
standing waves combine in a nearly linear way to produce an effect of
constructive and destructive interference, causing spots and other
features to drift towards a preferred alignment. A similar explanation
exists for spot interactions in the reaction-diffusion system of
\cite{schenk1998}. Analytical research is needed to establish a basis
for this theory for the Gray-Scott model equations, or another
explanation that can account for the observed phenomena. Inasmuch as
similar effects are observed in the 1-dimensional system (see Mazin
et. al \cite{mazin1996} and my own work \cite{munafo2010slides}),
analytical work can probably begin with 1-dimensional systems.

There is much opportunity for further research by numerical
simulation, including rigorous statistical analysis to establish the
degree of stability of the patterns in response to varying types and
levels of perturbation. The majority of the present results are at a
single point in the parameter space; there is a likelihood that
changing $F$ and $\sigma=D_u/D_v$ will yield new discoveries.

The superficial similarity (spot shape and ``halos'' and target
patterns) and behavior (spots condensing into a bound multi-spot state
similar to ``molecules'') in gas plas\-ma experiments, combined with the
successful modeling (e.g. by \cite{liehr2000}) of these experiments
by a reaction-diffusion system in numerical simulation, strongly suggest
research to identify more connections between the present work
and the gas discharge systems.

Concerning the moving stable structures, there are several well-studied
reac\-tion-dif\-fu\-sion system models, including the Brusselator
model with diffusion, the three-com\-po\-nent sys\-tem of \cite{liehr2000},
and the Oregonator model of the B-Z reaction \cite{yang2002}; all
sharing many or most of the features of the Gray-Scott system
(stripes, spots, multiple homogeneous stable states, mixed-mode
patterns, ability to produce Turing patterns, and varying types of
movement and interaction at various parameter values). Because of
these many shared traits, it seems likely that some of these other
systems can also produce stable moving localized patterns and other
complex phenomena like those described in the present paper. If such
phenomena exist, they likely involve parameter values that have not
been thoroughly explored, or may be confined to a very narrow region
of the parameter space.

Finally, this author notes that the great diversity of patterns and
types of interaction displayed by the Gray-Scott system at these
specific parameter values, combined with the inherent stability of
these patterns, clearly places this system in Wolfram's class 4
(complex localized structures, sometimes long-lived)
\cite{wolfram1984} that characterizes certain discrete cellular
automata that have been shown to be capable of universal computation.
Constructing large systems of interacting patterns presents several
challenges for long-term stability, because most interactions cause
each interacting part to shift. However, a starting pattern containing
an infinite array of elements, each of which is only used a small
number of times, would appear promising.

\section*{Acknowledgments}

The author acknowledges Jeff McGough for early feedback; thanks also
to Joseph Fineman for his copy-editing experience; and to Jonathan
Lidbeck for his 3-D Gray-Scott Java applet \cite{lidbeck2007}.


\end{document}